\documentclass[aps,psfig,prc,preprint,epsfig,showpacs]{revtex4}
\usepackage{graphicx}

\begin{document}

\title{p+\( ^{4,6,8} \)He elastic scattering at intermediate energies}
\author{E. Baldini-Neto, B. V. Carlson, R. A. Rego}
\affiliation{Departamento de F\'{\i}sica, Instituto Tecnol\'{o}gico de Aeron\'{a}utica, 
Centro T\'{e}cnico Aeroespacial, \\ 12228-900 S\~{a}o Jos\'{e} dos Campos, S\~{a}o
Paulo, Brazil}
\author{M. S. Hussein}
\affiliation{Nuclear Theory and Elementary Particle Phenomenology Group \\ 
Instituto de F\'{\i}sica, Universidade de S\~{a}o Paulo\\ 05508-900, S\~{a}o Paulo, Brazil}

\date{\today}

\begin{abstract}
Using a relativistic nuclear optical potential consisting of a Lorentz
scalar, \( V_{s} \), and the time-like component of a four-vector
potential, \( V_{0} \), we calculate elastic scattering differential
cross sections and polarizations for \( p+^{4} \)He at intermediate
energies for which experimental data are available. We also calculate
the differential cross sections and analyzing powers for \( p+^{6,8} \)He
at intermediate energies and compare with the few available experimental data.
\end{abstract}
\maketitle
\section{Introduction}
\noindent

The scattering of medium-energy nucleons from nuclei can provide
information about both nuclear structure and the \( NN \)-interaction. 
In particular, the study of proton-nucleus scattering at intermediate energies is
a useful method for determining accurate nuclear matter distributions in stable
nuclei \cite{Al 78}. Recently, it has also been used extensively to determine the extension
of the nuclear matter density of exotic nuclei. In particular, in
a recent set of experiments, inverse kinematics was used to study
the elastic scattering of \( p+^{4,6,8} \)He and determine their matter
densities \cite{Al 97, Eg 01}.

Elastic scattering at intermediate energies can be calculated theoretically using
either a non-relativistic \cite{Ak 98, Am 02} or a relativistic optical \cite{Ar 79, Ar 81} 
model, usually with about the same good results. An advantage of a relativistic optical
model is that its two potentials simultaneously determine both the
central and spin-orbit interactions. When the relativistic impulse
approximation (RIA) is used \cite{Mc 83}, these interactions may be determined
in terms of the corresponding scalar and vector nuclear densities \cite{Se 86}.

In this work we use a relativistic nuclear optical potential constructed from a
Lorentz scalar, \( V_{s} \), and the time-like component of the four-vector
potential, \( V_{0} \), to calculate elastic scattering angular distributions
and analyzing powers for \( p+^{4} \)He at intermediate energies
and compare two different fits to the \( p+^{4} \)He data and also to the RIA results.
We also calculate angular distributions and analyzing powers of the elastic scattering 
of \( p+^{6,8} \)He and compare an extension of the more physical of our two fits with the RIA 
and the few available experimental data.

\section{The Dirac Optical Potential}
\noindent

In relativistic optical model analyses of intermediate energy
scattering \cite{Ar 79}, the Dirac equation is usually used in the form,
\begin{equation}
\label{eq1}
\{\vec{\alpha }\cdot \vec{p}+\beta [m+V_{s}(r)]+[V_{0}(r)+V_{c}(r)]\}\Psi (\vec{r})=
E\Psi (\vec{r}),
\end{equation}
where \( V_{s} \) is an attractive Lorentz scalar potential, \( V_{0} \)
is the repulsive time-like component of a four-vector potential and
\( V_{c} \) is the Coulomb potential. The choice of the potentials
is motivated by meson exchange considerations and simplicity.
The simplest meson exchange interaction possessing a certain justification on physical grounds,
which is also capable of providing nuclear saturation properties, takes into account the exchange
of an attractive intermediate range isoscalar scalar meson and a repulsive shorter range isoscalar
vector meson \cite{Se 86, Er 74, Br 76, Ue 78}. 
The two corresponding potentials \( V_{s} \) and \( V_{0} \) play an essential
role in the description of both elastic scattering and polarization
data at intermediate energies since their sum is the principal contribution
to the central potential while their difference determines the spin-orbit
interaction \cite{Fu 36}. The scalar, \( V_{s} \), and vector, \( V_{0} \),
optical potentials used in the analyses here can be written as
\begin{eqnarray}
 &  & V_{s}=U_{s}f_{U_{s}}(r)+\imath W_{s}f_{W_{s}}(r)\nonumber \label{eq2} \\
 &  & V_{0}=U_{0}f_{U_{0}}(r)+\imath W_{0}f_{W_{0}}(r)
\end{eqnarray}
with each of the potentials possessing both real and imaginary parts with possibly
different radial dependences.

Based on nuclear structure calculations, we take the radial dependence \( f(r) \), for the case 
of $p+^{4}$He, to have a Gaussian form 
\begin{eqnarray}
f(r) & = & exp[-r^{2}/r^{2}_{0}].\label{eq3} 
\end{eqnarray}

\section{Results}
\subsection{\protect\( p+^{4}\protect \)He elastic scattering}
\noindent

\begin{figure}[htb]
\centerline{\includegraphics[angle=-90.0,width=10.00cm]{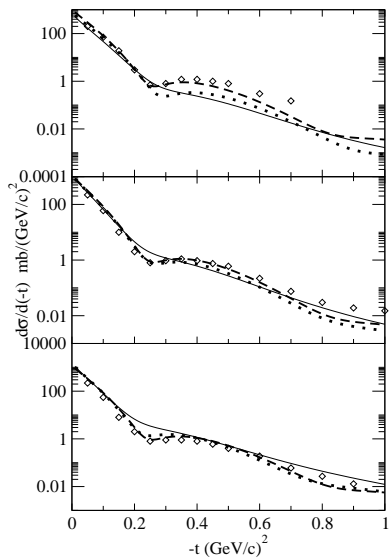}}
\caption{\label{fig1}Differential cross section of $p+^{4}$He, as a function of the 
momentum transfer, calculated for $E_{lab}=0.561$, $0.800$ and $1.029$ GeV.
The solid lines represents the RIA calculation while the dashed and dotted ones stand for 
the first and second fits. The experimental data are labeled with diamonds.} 
\end{figure}

\begin{figure}[htb]
\centerline{\includegraphics[angle=-90.0,width=10.00cm]{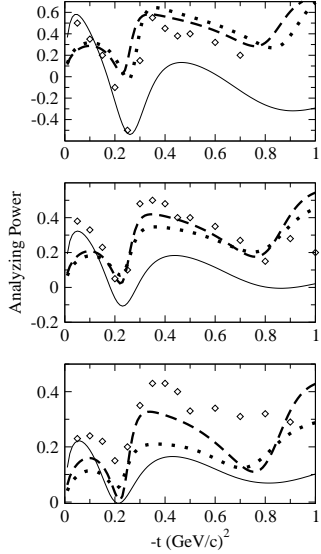}}
\caption{\label{fig2}Polarization of $p+^{4}$He, as a function of the momentum transfer,
 calculated for $E_{lab}=0.561$, $0.800$, $1.029$ GeV.
The solid lines represents the RIA approximation while the dashed and dotted ones stand for 
the first and second fits. The experimental data are labeled with diamonds.}
\end{figure}

Relativistic Hartree calculations of the \( ^{4} \)He nucleus yield
almost identical scalar and vector densities but an rms radius
about 30\% larger than the experimental one \cite{Br 00}. To obtain a physically
reasonable RIA potential for this system, we use equal scalar and vector
matter densities having the rms radius of $1.49$ fm found in Ref. \cite{Al 97}
together with the relativistic Love-Franey NN t-matrix of Ref. \cite{Ho 85}.
The results, represented by full lines in Figs. (\ref{fig1})-(\ref{fig2}),
show that the RIA agrees well with the experimental data at low momentum
transfer but deviates substantially from the experimental results as
the momentum transfer increases. We also note that the RIA angular
distributions do not reproduce the oscilatory structure seen in the experimental data, 
decreasing monotonically instead. The polarizations are not reproduced either, except at very
low values of the momentum transfer. 
Due to these differences, we have tried to fit the experimental data of
Ref. \cite{Ar 79} using a Dirac optical potential of the form given
in equations (\ref{eq2})-(\ref{eq3}).

In a first attempt, labeled with the dashed lines in the figures, the
parameters were left free to vary so as to obtain the simultaneous
best fit to the experimental angular distributions and polarizations at
the three values of the laboratory energy, $E_{lab}$.
The same geometrical parameters were used at
the three energies, while the strengths were assumed to vary
linearly with the laboratory energy as $V=V_{0}+V_{1}*E_{lab}$,
a common parametrization of optical model strengths. The best fit parameters
are given in Table \ref{table1}. 
As can be seen in the figures, we indeed obtain
a fairly reasonable fit to the angular distribution and polarization
at all three energies. However, an interpretation of the imaginary
part of the potential in the context of a RIA potential
would require a $NN$ total cross section about 2.5 times the physical
one, which makes the fit unsatisfactory on physical grounds. The radii
of the imaginary potentials are in agreement with the matter radius
of Ref. \cite{Al 97}, while the radii of the real potentials are
found to be about \( 10\% \) bigger. These also disagree with what one would expect from
a RIA potential, which usually yields real radii close to the matter radius and
imaginary ones slightly larger.
 
To obtain a more physically reasonable fit to the \( p+^{4} \)He
data, we reduced the imaginary strengths of the first fit
by a factor of 3, reset the radii to the value of \( r_{0}=1.22 \)
fm, corresponding to an rms radius of $1.49$ fm \cite{Al 97}, and then let 
the parameters vary freely once again. The results,
labeled with dotted lines in the figures, present slightly poorer
agreement with the experimental data, but continue to reproduce the
oscillations in the angular distributions that are not obtained in the RIA
calculation (full lines). The imaginary potentials of the second fit are
consistent with those of the RIA potential, corresponding to a physically
reasonable $NN$ total cross section. The radii are also closer
to those of a RIA potential. 

\begin{table}
\caption{\label{table1}Optical parameters for fits to the $p+^{4}$He experimental data.}
\begin{ruledtabular}
\begin{tabular}{crrr}
             & $V_{0}$(MeV) &  $r_{0}$(fm)  &  $V_{1}$  \\   \hline
$U_{s}^{1}$  & $-325.50$  &  $1.354$    & $ 0.1051$ \\
$W_{s}^{1}$  & $ 208.65$  &  $1.209$    & $ 0.0147$ \\
$U_{0}^{1}$  & $ 271.84$  &  $1.346$    & $-0.0160$ \\
$W_{0}^{1}$  & $-300.00$  &  $1.186$    & $-0.0538$ \\   \hline
$U_{s}^{2}$  & $-325.50$  &  $1.309$    & $ 0.1162$ \\
$W_{s}^{2}$  & $  79.65$  &  $1.295$    & $ 0.0653$ \\
$U_{0}^{2}$  & $ 271.84$  &  $1.252$    & $-0.0540$ \\
$W_{0}^{2}$  & $-100.00$  &  $1.257$    & $-0.1658$ \\ 
\end{tabular}
\end{ruledtabular}
\end{table}

\newpage

\subsection{\protect\( p+^{6,8}\protect \)He elastic scattering}
\noindent

\begin{figure}[htb]
\centerline{\includegraphics[angle=-90.0,width=10.0cm]{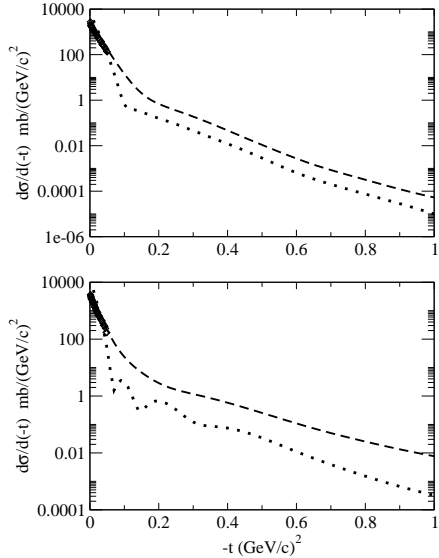}}
\caption{\label{fig3}Differential cross sections of $p+^{6,8}$He at 
$E_{Lab}=0.721$, $0.678$ GeV, respectively. The dashed line corresponds
to the RIA approximation while the dotted one represents the second fit. 
The experimental data are labeled with diamonds.}
\end{figure}

\begin{figure}[htb]
\centerline{\includegraphics[angle=-90.0,width=10.0cm]{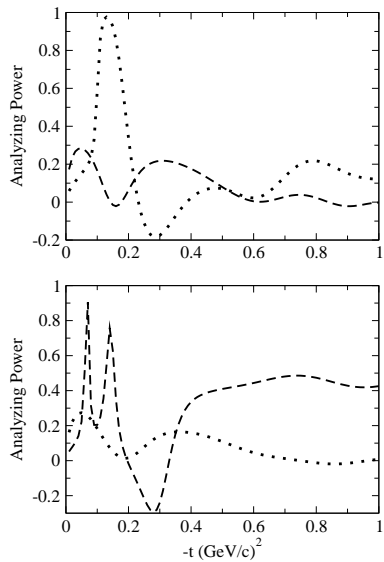}}
\caption{\label{fig4}Polarization of $p+^{6,8}$He at 
$E_{Lab}=0.721$, $0.678$ GeV respectively. The dashed line corresponds
to the RIA approximation while the dotted one represents the second fit.}
\end{figure}

We next considered the elastic scattering of $p+^{6,8}$He at
laboratory energies of $E_{lab}=0.721$ GeV and $E_{lab}=0.678$ GeV, respectively.
To perform the RIA calculation for \( ^{6,8} \)He, we followed the
same procedure used for \( ^{4} \)He. We generated equal scalar and
vector matter densities using the appropriate Gaussian harmonic oscilator (HO)
geometrical parameters of Ref. \cite{Al 97}, together with the relativistic Love-Franey
NN t-matrix of Ref. \cite{Ho 85}. The results of the RIA calculation
in Figs. (\ref{fig3}) and (\ref{fig4}) agree well with the experimental
data in the range over which the data exist. We emphazise that the RIA also describes the
$p+^{4}$He data well in this range of momentum transfers but, at larger momentum transfers,
deviates substantially from the data in form and absolute values.

For the cases of $p+^{6,8}$He, the data is insufficient to attempt a fit. 
We have thus attempted to extend our fits to the \( ^{4} \)He data by interpreting
them in the context of a simple RIA calculation. To do so, we extract
the effective strengths, \( \tilde{V}_{0i}=\frac{1}{4}(U_{i},W_{i})[\pi r_{0,i}^{2}]^{3/2} \),
that, when multiplied by the \( ^{4} \)He density, would yield the
potentials obtained in our fits to the $p+^{4}$He data. We have multiplied these
strengths by the appropriate \( ^{6,8} \)He densities, obtained using
the Gaussian HO parametrization of Ref. \cite{Al 97} to obtain effective 
Dirac potentials for these systems,

\begin{eqnarray}
\label{eq4}
 V_{i}(r)&=&\tilde{V}_{0i}\left( \frac{4}{(\pi r_{0c}^{2})^{3/2}}\exp\left[-r^{2}/r_{0c}^{2}\right]+
\right.  \nonumber  \\
    && \left.\zeta \frac{2}{3}\frac{1}{(\pi r_{0v}^{2})^{3/2}}\frac{r^{2}}{r_{0v}^{2}}
\exp\left[-r^{2}/r_{0v}^{2}\right]\right),  
\end{eqnarray}
where \( \zeta =2,4 \) for \( ^{6} \)He and \( ^{8} \)He, respectively.
The radii were modified slightly to take into account the differences
between the experimental \( ^{4} \)He radius of Ref. \cite{Al 97} and those 
obtained in the fits. 

Our calculations for these two nuclei do not reproduce the experimental data
as well as the RIA calculation does, as can be seen in Figs. (\ref{fig3}) and (\ref{fig4}), which
leads us to conclude that the parameters obtained in the $p+^{4}$He
fit cannot be so easily extended to the \( p+^{6,8} \)He elastic
scattering. However, we feel that they can transmit an idea of the
discrepancies with an RIA calculation that the experimental values
could show. In Figure (\ref{fig4}) we show the
polarizations expected for proton scattering from these nuclei, for
both the RIA and our extension of the \( ^{4} \)He fits. The differences
between the two are again quite large. However, no data exist in this case. 
By comparision with the results for  \( ^{4} \)He it is difficult to say what one could expect of
the experimental angular distribution and polarization in \( p+^{6,8} \)He scattering
at larger values of the momentum transfer. 


\newpage
\section{Conclusion}
\noindent

We have calculated $p+^{4,6,8}$He elastic scattering differential cross sections 
and polarizations using the relativistic impulse approximation and an adjusted 
Dirac optical potential. 
We have shown that the RIA results obtained using the parameters of Ref. \cite{Al 97} describe
all three systems well at low values of the momentum transfer. However, in the case of
$p+^{4}$He, for which more data exist, the RIA deviates substantially from the data at higher 
momentum transfers. As an alternative to the RIA we have ajusted the parameters of a Dirac 
optical potential and then attempted to extend the parameters obtained to the cases of 
$p+^{6,8}$He, but with unsatisfactory results. Based on our analyses it is difficult to 
say what one could expect at higher values of transferred momenta in the cases of $p+^{6,8}$He. 
More data are needed to elucidate this situation. 
  
In closing, we should mention that a possible reason for the failure of RIA in describing
the $p+^{4}$He data is that it has been pushed beyond the limit of its applicability in the simple 
manner in which it is used here. The decomposition of the effective Dirac potential into a scalar 
potential and the fourth component of vector is valid in the target rest frame. 
The scattering calculation, however, is performed in the CM frame and the boost that carries one 
frame to the other will convert the vector fourth component potential into a full vector potential. 
For proton scattering on a system such as $^{40}$Ca the vector components introduced are small 
and can be neglected. For extremely light systems, such as those studied here, this is not the 
case. We plan to examine the importance of this effect in the future.


\begin{references}

\bibitem{Al 78}G. D. Alkhazov \textit{et al.}, \textit{Phys. Rep.} \textbf{C 42}, (1978), 89.
\bibitem{Al 97}G. D. Alkhazov \textit{et al.}, \textit{Phys. Rev. Lett} \textbf{78}, (1997),
2313.
\bibitem{Eg 01}P. Egelhof, \textit{Progr. Part. Nucl. Phys.} \textbf{46}, (2001), 307.
\bibitem{Ak 98}J. S. Al-Khalili and J. A. Tostevin, \textit{Phys. Rev.} 
\textbf{C 57}, (1998), 1846.
\bibitem{Am 02}K. Amos, S. Karataglidis and P. K. Deb, \textit{nucl-th/0202081}, (2002).
\bibitem{Ar 79}L. G. Arnold, B. C. Clark and R. L. Mercer, \textit{Phys. Rev.} \textbf{C 19},
(1979), 917.
\bibitem{Ar 81}L. G. Arnold, B. C. Clark, R. L. Mercer and P. Schwandt, \textit{Phys. Rev.}
\textbf{C 23}, (1981), 1949.
\bibitem{Mc 83}J. A. McNeil, J. R. Shepard and S. J. Wallace, \textit{ Phys. Rev. Lett.} 
\textbf{50}, (1983), 1439, 1443.
\bibitem{Se 86}B. D. Serot and J. D. Walecka, \textit{Advances in Nuclear Physics}, ``The
Relativistic Many-Problem``, Plenun Press - New York (1986).
\bibitem{Er 74}K. Erkelenz, \textit{Phys. Rep.} \textbf{13 C} (1974), 194.
\bibitem{Br 76}G. Brown and D. Jackson, \textit{The nucleon-nucleon interaction}, 
(North-Holland-Amsterdam), 1976.
\bibitem{Ue 78}T. Ueda, F. E. Riewe and A. E. S. Grenn, \textit{Phys. Rev.} \textbf{C 17},
(1978), 1763.
\bibitem{Fu 36}W. H. Furry, \textit{Phys. Rev.} \textbf{50}, (1936), 784. 
\bibitem{Br 00}B. V. Carlson and D. Hirata, \textit{Phys. Rev.} \textbf{C 62}, (2000), 054310. 
\bibitem{Ho 85}C. J. Horowitz, \textit{Phys. Rev.} \textbf{C 31}, (1985), 1340. 

\end{references}
\end{document}